\documentclass[conference]{IEEEtran}
\IEEEoverridecommandlockouts
\usepackage{cite}
\usepackage{amsmath,amssymb,amsfonts}
\usepackage{algorithmic}
\usepackage{graphicx}
\usepackage{textcomp}
\usepackage{xcolor}
\def\BibTeX{{\rm B\kern-.05em{\sc i\kern-.025em b}\kern-.08em
    T\kern-.1667em\lower.7ex\hbox{E}\kern-.125emX}}
\begin{document}

\title{Blockchain based solution design for Energy Exchange Platform\\
}

\author{
\IEEEauthorblockN{Atharv Bhadange}
\IEEEauthorblockA{\textit{dept. computer engineering} \\
\textit{Pune Institute Of Computer Technology}\\
Pune, India \\
bhadange.atharv@gmail.com}
\and
\IEEEauthorblockN{Rohan Doshi}
\IEEEauthorblockA{\textit{dept. computer engineering} \\
\textit{Pune Institute Of Computer Technology}\\
Pune, India \\
rohan.doshi02@gmail.com}
\and
\IEEEauthorblockN{Tanmay Karmarkar}
\IEEEauthorblockA{\textit{dept. computer engineering} \\
\textit{Pune Institute Of Computer Technology}\\
Pune, India \\
tanmaykarmarkar49@gmail.com}
\and
\IEEEauthorblockN{Snehal Shintre}
\IEEEauthorblockA{\textit{Assistant Professor} \\
\textit{Pune Institute Of Computer Technology}\\
Pune, India \\
spshintre@pict.edu}
}

\maketitle

\begin{abstract}
It is observed that users have higher requirements for fairness, transparency, and privacy of transactions of energy exchanges that occur across platforms like Indian Energy Exchange (IEX) and Power Exchange India Limited (PXIL). As a decentralized distributed accounting system, blockchain is characterized by traceability, security, credibility, and non-tampering of transactions, which can meet the needs of integrated energy and multi-energy transactions. Based on the research on the application of blockchain technology in the field of integrated energy services, this solution proposes an integrated energy trading process based on smart contracts and explores the application of blockchain technology in integrated energy services.
\end{abstract}
\vspace{12pt}

\section{Introduction}
Blockchain technology can be used in ways such that we can use blockchain to verify the data structure and data storage and use distributed node consensus algorithm to generate and update the data, use of cryptography way to ensure the security of data transmission and access, use automated script code intelligent contracts to program and operate data of a new kind of distributed infrastructure and computing paradigm. Features such as "Honesty" and "Transparency" of the blockchain are guaranteed which lay the foundation stone for creating trust in the blockchain. The huge application scenarios of blockchain are based on blockchain to solve the problem of information asymmetry and realize cooperative trust and concerted action among multiple subjects.

If a centralized decision-making method were to be proposed it would make the operation cost of transaction centers become higher and the time consumption increase. If hackers attack the transaction center, the security of the transaction and the privacy of the participants cannot be guaranteed. The proposed blockchain solution leads to the process of decentralized decision-making in transactions which lacks a central oversight mechanism on how to verify the identities of both parties to ensure transaction security.

Cryptography can be used in the process of trade disputes by blockchain technology, which proposes a decentralized payment system. By effectively solving the problem of information asymmetry in the trading process, we are confident that blockchain technology will not manipulate the characteristics of low-cost, information-integrated energy trading applications. The technical characteristics of blockchain, such as security and reliability, openness and transparency, and collective maintenance, can meet the needs of integrated energy market transactions.

Blockchain can be compared to distributed database technology. The chain structure of data blocks allows you to manage a continuously growing and unmanipulated data set. Features of blockchain include decentralization, collective maintenance, smart contracts, security, and trust. As a typical programmable technology, smart contracts can describe the transaction execution process in an auto-programmable language and execute pre-embedded commands. Code for automation and ensuring integrity of transactional execution.

\section{Motivation}
Over the past few years. In order to reduce the central control authority over documents, and transactions, and also prevent any changes to them, blockchain is the ideal technology. While being anonymous, blockchain features non-modifiable transactions, which are useful for facilitating transactions between many government organizations and private institutions. Growth of Ethereum and Solona to write smart contacts. The concept of a smart contract refers to a dynamic, multi-recognized program that runs on the blockchain in response to events and can be automated to process assets automatically in accordance with certain conditions. Smart contracts' largest benefit is replacing artificial arbitration and agreement enforcement with program algorithms. Considering all the features of blockchain, it is well suited for the problem of scalable, decentralized, and a secured systems of energy exchanges.

\section{Literature Survey}

\subsection{Research on Design and Application of Power Dispatch Based on Blockchain}
Power dispatch urgently needs to improve the level of instant information sharing and service management capabilities. Through blockchain technology, power dispatch can realize the credible sharing and traceability of key data. It can reduce the impact of the misoperation of the security and stability control device on the power grid. The power system is a huge and complex system. Many power plants produce electric energy and supply power to users through transmission, transformation, distribution, and power supply networks. At present, with the operation of UHV AC/DC hybrid power grids, large-scale consumption of renewable energy, and rapid development of electricity market operations, the scale of power grids is rapidly expanding, and power dispatch structure is becoming increasingly complex, which puts forward higher requirements for integrated operation of large power grids. In order to standardize the grid-connected operation management of power generation companies and reasonably evaluate the quality of their auxiliary services, the energy regulatory agency will formulate grid operation management specifications and conduct assessments on power plants and other relevant entities to promote safe and stable operation of the power grid.\cite{b1}.

\subsection{Application of Blockchain in Intelligent Power Exchange}
This paper suggests the use of blockchain for power transaction management to redesign business processes. It introduces a "chain-on-chain" idea to plan the data-chaining scheme and control issues encountered during data exchange to improve the intelligent power data quality. It defines a DTBP (Decentralized Transaction credit model based on Blockchain P2P) using some classic node trust model indicators based on an aggregation of transaction history and recommendation degree.\cite{b2}.

\subsection{Research on blockchain in energy internet}
Blockchain has gradually become the underlying technology for the future value of the Internet. In order to solve the problem of large-scale promotion and application of new energy, this paper puts forward the concept of energy Internet. The Energy Internet takes the power system as the center, integration of smart grid, Internet, big data, cloud computing and other cutting edge technology, advanced technology, comprehensive management, implementation of energy exchange of complementary, fusion energy systems into the next generation.
Distributed energy and microgrid will become an important part of the energy Internet," the consumer is the producer", emphasizing equal energy Advances in Computer Science Research, Volume 74 754 sharing among individuals.
Characteristics Of Energy Internet:
1. Accurate measurement	
2. Ubiquitous interactions
3. Autonomous control
4. Optimization decision
5. Wide area coordination
\cite{b3}.

\subsection{A Blockchain-based Trading Matching Scheme in Energy Internet}
To solve the problem of distributed energy exchange trading in the energy Internet, the paper proposes a solution based on the electrical energy trading system BC-TMS Blockchain-based trading matching system, considering the security concerns that are present. The paper provides a solution to achieve accurate matching of trading users according to user needs. The goal is to maximize satisfaction to users while having maximum benefits. It does so by introducing virtual agent nodes to assist buyers and sellers. Using blockchain we can hide the users' private information and identify malicious users in the system. It cannot cause the market to lose fairness. The design goals of the paper are broadly categorized into decentralization, trading mechanism, privacy-preserving, automatic execution, and high efficiency. The given trading algorithm consists of trading nodes and agent nodes ensuring maximum buyer and seller satisfaction. Using blockchain 3.0, we can confirm, measure, and store digital assets so that assets can be tracked, controlled, and traded on the blockchain\cite{b4}.

\subsection{Distributed Energy Transaction Mechanism Design Based on Smart Contract}
The existing power market is designed according to the government, central power supply, public services, and power market. The traditional energy transaction is carried out in the following four steps: auditing, bidding, clearing, and settlement. The paper proposes a P2P system that combines the energy produced, consumers, producers, energy storage, and energy exchange in a microgrid, and a P2P transaction mode is introduced. The paper proposes to set up an intelligent contract to realize the unsupervised electric auction. The proposed solution in the paper carries out the whole process in 5 steps:
1. Requirement Publishing: Producers publish their excess energy on the platform.
2. Qualification Auditing: The platform will audit the trader’s qualifications.
3. Transaction Confirmation: Consumers will bid. The platform will verify the capability of consumers and producers and process the transaction.
4. Transaction Execution: Consumers finalize the transaction. 
5. Transaction Settlement: Bills are automatically settled.
\cite{b5}.

\begin{figure*}
  \centering \includegraphics{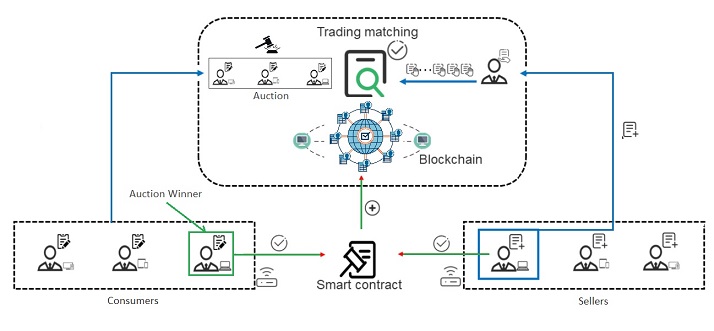}
  \caption{Auction-Based energy trading model}
\end{figure*}

\subsection{Smart contracts in energy systems: A systematic review of fundamental approaches and implementations}
This paper proposes a Decentralised Autonomous Area Agent (D3 A) Market Model, an open energy exchange engine to model, simulate and operate energy trading markets in local communities. The energy exchange can be operated by unique producers or multiple agents using smart contracts, to define the energy trading and matching between the customers. A Power Ledger deploys a dual-token ecosystem with a PoA consensus mechanism to decrease energy consumption, limit double-spend tokens, and control access to the chain. The Power Ledger platform allows the producers to manage a microgrid with a real-time energy market and traceable renewable energy certificates. Most of the smart contracts for P2P energy trading implement a function that receives and saves the bids or requirements of end-users. Bids can include the amount of energy, the time at which the energy is needed or available, the price that is desired to buy or sell, the proposed quantity of energy, and finally, also the power.
\cite{b6}.

\section{System Description}
\subsection{Blockchain}
The simplest way to describe Blockchain is by saying that it is a distributed ledger. In simple words, you can say it’s a distributed database or a decentralized database. A ledger is a “book or collection of accounts where transactions are recorded”. Blockchain is a distributed database where the record of all the changes and transactions is shared among the people owning a copy of the blockchain. Not only is the blockchain secured and distributed, but the records also cannot be tampered with. Blockchain features a consensus that makes unbiased decisions. Also, the signature of all the records is verified before adding to the transaction list while the information behind it is kept hidden byways of the cryptographic algorithms.
\begin{enumerate}
    \item Decentralization:
There is no central data storage node in the blockchain. All the nodes store all the data which helps in accomplishing complete decentralization. Decentralization can also be adopted by maintaining a weakly managed central node that does not pose a threat to security.
    \item Counterfeit-Resistant:
Security is provided by storing the signatures of both parties and the signs cannot be forged.
    \item Verifiability:
Electronic money can be traced back to its origin and the input in any transaction is the output of the previous transaction. The transaction amount, verification of source, and correctness of it can all be traced back to its root which makes every transaction reliable.
    \item Anonymity:
The transactions over the blockchain are visible to each and every user that is part of the blockchain. The identity of transaction nodes is masked by a series of Hash Operations which results in the security and anonymity of participating nodes.
 \end{enumerate}
\vspace{10pt}

 \begin{figure}[htbp]
\centering
{\includegraphics[width = 70mm]{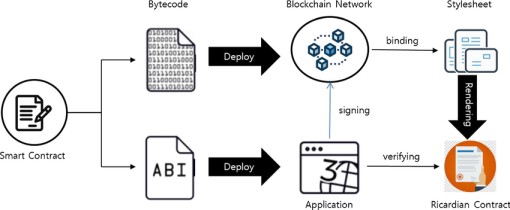}}
\caption{Smart contract}
\label{fig}
\end{figure}

\vspace{10pt}

\subsection{Smart Contracts}
A smart contract is basically like a trigger that helps in automating transactions. When predetermined conditions are met a smart contract is automatically activated. The contract contains (if/else … then) ladder code contained the blockchain. They are useful in transactions as there is no time lag present between the transaction and the outcome of the transaction. Once the contract is successfully executed the blockchain is updated and the data is secured. Only the parties involved in the transaction can view this transaction. Cryptography can make sure that assets are safe and sound even if someone breaks the encryption, they would have to modify all the blocks that come after the block that has been modified. The monetary budget is curbed as smart contracts eliminate the presence of intermediates. All parties which are involved in the agreement are distributed funds agreed upon in the agreement as smart contracts support multi-signature accounts.

\begin{figure}[htbp]
\centering
{\includegraphics[width = 70mm]{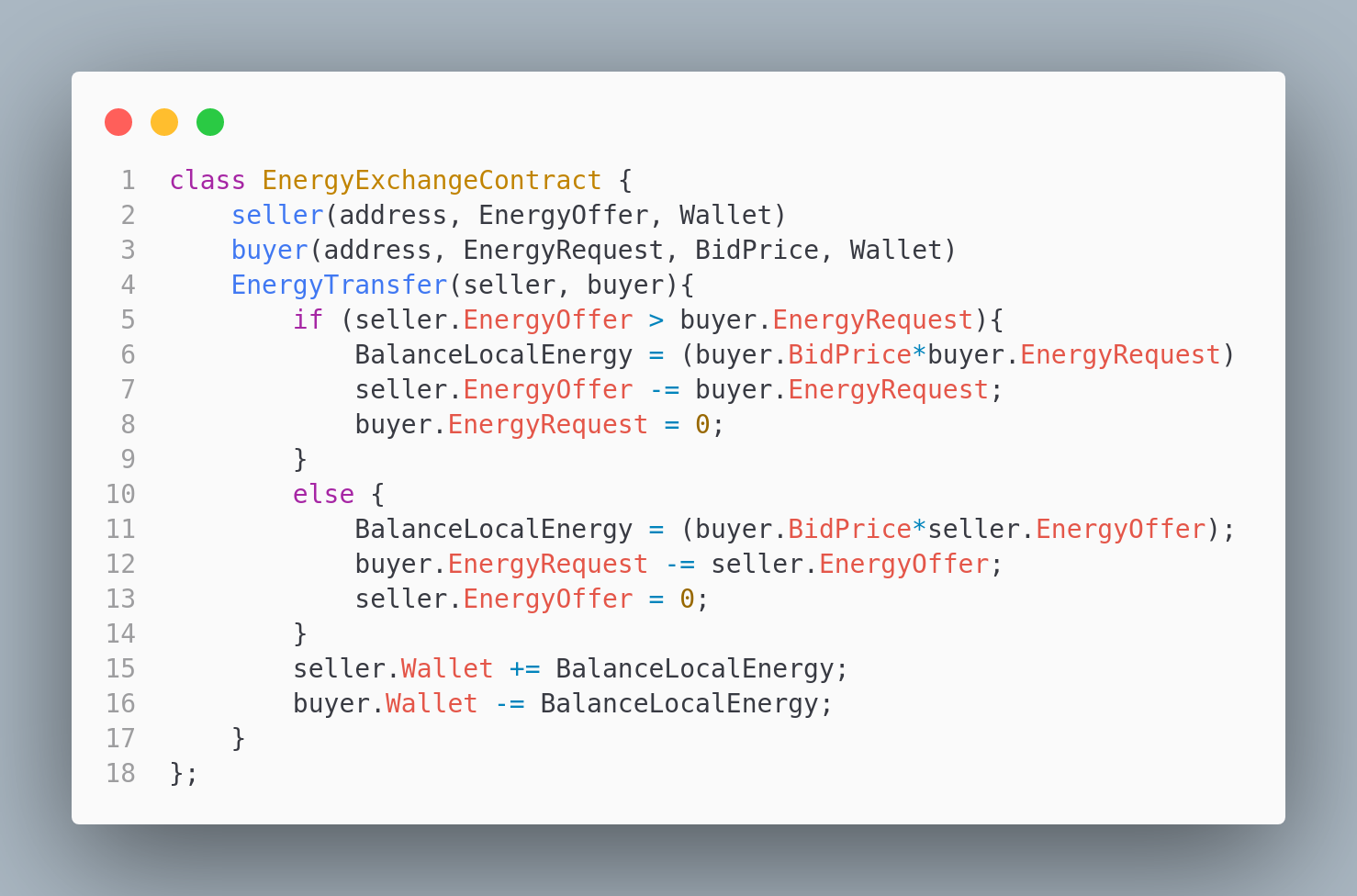}}
\caption{Algorithm for energy transfer contract}
\label{fig}
\end{figure}

\vspace{10pt}
The system provides a design and implements a decentralized trading model in the context of large-scale access to distributed energy. The mechanism needs to be able to match supply and demand information effectively while achieving high efficiency and low cost of trading. The system designs a trade matching mechanism to match buyers and sellers to complete the transaction. The proposed solution will be having 2 types of nodes (Buyer nodes and Seller nodes). In order to achieve maximum satisfaction for both the seller nodes and buyer nodes, we are proposing a solution that consists of trade matching with the help of an auction system. The seller node will have a base auction cost for the electricity value that it will be accounting for.  The auction will then be open for a particular time period during which the buyers would bid in the election process. Here, the buyer with the maximum auction fee will be given the contract and a transaction between the seller and buyer would be completed and the result will be reflected on the blockchain. Other buyers would be considered for the remaining sellers for other auctions that are occurring concurrently. Situations to consider here would be that the seller would require gas fees to initiate an auction. The auction will be available for a fixed amount of time. If there are no interested buyers and the time for auction expires no contract on the blockchain would be published and the contract would be discarded. The auction between the buyer and seller with a base auction price would result in the maximum satisfaction of the buyer and seller. Another possibility is that the buyer would not be interested in providing direct ether to the seller but a bond that could be issued by the buyer to the seller. The bond could be provided through a non-fungible token (NTF) and can be treated as an asset. The NFT could be further traded having a base value of the transaction.

\section{Conclusion}
The paper presents a blockchain-based decentralized Energy Exchange Platform solution where the Buyers and Sellers are matched to complete the transaction. The solution consists of an auction system and to achieve the optimum solution the electricity will have a base price. The auction will be closed after the designated amount of time and the winner of the auction will be declared and the transaction will be completed. The transaction will be pushed onto the blockchain.

\vspace{12pt}

\end{document}